\documentclass[11pt]{article}
\usepackage[ margin=1in]{geometry}
\usepackage[affil-it]{authblk}
\usepackage{setspace}
\usepackage{tocloft}

\usepackage[sort&compress,numbers, super]{natbib}
\usepackage[font=footnotesize]{caption}
\usepackage{graphicx}% Include figure files
\usepackage{amsmath,amssymb,bm, bbold}% bold math
\usepackage{xr}
\usepackage{color}
\usepackage[normalem]{ulem}
\usepackage{wasysym}
\usepackage{siunitx}
\usepackage{physics}
\usepackage{lineno}
\usepackage{gensymb}
\usepackage[T1]{fontenc}
\usepackage{booktabs}
\usepackage[section]{placeins}
\usepackage{textcomp}
\usepackage[colorlinks=true, linkcolor=black, urlcolor=blue, citecolor=blue]{hyperref}
\usepackage{verbatim}
\usepackage{blindtext}
\usepackage{comment}

\usepackage{upgreek}
\usepackage{textgreek}
\usepackage{romannum}

\makeatletter

\newcommand{\Rmnum}[1]{\expandafter\@slowromancap\romannumeral #1@}
\makeatother

\definecolor{mygreen1}{rgb}{0, 0.5, 0.7}



\begin{document}
	
\title{\vspace{-1cm}\textbf{Highly nonlinear dipolar exciton-polaritons in bilayer MoS$_2$}}

\author[1,*]{Biswajit Datta}
\affil[1]{Department of Physics, City College of New York, New York, NY, USA}

\author[1,2]{Mandeep Khatoniar}
\affil[2]{Department of Physics, Graduate Center of the City University of New York (CUNY), New York, NY, USA}

\author[1,2]{Prathmesh Deshmukh}

\author[3]{Félix Thouin}
\affil[3]{Department of Engineering Physics, École Polytechnique de Montréal, Montréal, Quebec, Canada}

\author[1,2]{Rezlind Bushati}

\author[4]{Simone De Liberato}
\affil[4]{School of Physics and Astronomy, University of Southampton, Southampton, UK}

\author[3]{Stephane Kena Cohen}

\author[1,2,*]{Vinod M. Menon}

%\affil[$\dagger$]{\textnormal{These two authors contributed equally to this work}}
\affil[*]{\textnormal{bdatta@ccny.cuny.edu, vmenon@ccny.cuny.edu}}

\date{}

\maketitle

\begin{abstract}
Realizing nonlinear optical response in the low photon density limit in solid-state
systems has been a long-standing challenge. Semiconductor microcavities in the
strong coupling regime hosting exciton-polaritons have emerged as attractive candidates in this context. However, the weak interaction between these quasiparticles has been a hurdle in this quest. Dipolar excitons provide an attractive strategy to overcome this limitation but are often hindered by their weak oscillator strength. The interlayer dipolar excitons in naturally occurring homobilayer MoS$_2$  alleviates this issue owing to their hybridization of interlayer charge transfer exciton and intralayer B exciton. Here we demonstrate
the formation of dipolar exciton polaritons in bilayer MoS$_2$  resulting in unprecedented nonlinear interaction strengths. A ten-fold increase in nonlinearity is observed for the interlayer dipolar excitons compared to the conventional A excitons. These highly nonlinear dipolar polaritons will likely be a frontrunner in the quest for solid-state quantum nonlinear devices.
\end{abstract}

\section*{Introduction}

Photons are becoming ubiquitous in emerging quantum technologies like quantum communication and metrology due to the ability to propagate long distances while being robust against decoherence~\cite{gisin2007quantum, giovannetti2004quantum}. To further extend the span of their utilities, platforms to achieve and implement strongly interacting photons in solid-state systems are highly desirable~\cite{chang2014quantum}. Conventional materials do not exhibit nonlinear response at power levels associated with single photons. In this context, remarkable advances have been made in cold atom systems to realize interacting photons at the single-particle level \cite{PhysRevLett.87.037901}. A strong contender for the generation of strongly interacting photons in the solid-state are exciton-polaritons formed via non-perturbative coupling of cavity photons with excitonic resonances. Although they can be modeled as non-interacting quantum fluids at low densities, beyond a critical density, saturation and short-range exchange interactions become significant and give rise to various phenomena like the appearance of spontaneous coherence, parametric down-conversion, and superfluidity \cite{Carusotto2013}. Strong spatial confinement along with Coulomb interactions give rise to even stronger correlations between polaritons that can no longer be described using a mean-field theory. Such interactions give rise to non-Poissonian statistics of laser transmission, dubbed as polariton blockade. Preliminary evidence of such non-classical correlation was recently observed in confined polariton systems in GaAs~\cite{Munoz-Matutano2019,Delteil2019} but a small ratio of interaction to dissipation rates resulted in only a weak violation of classical correlations.\\

Transition metal dichalcogenides (TMDCs) have garnered much attention for their exceptional optoelectronic properties and have been used to demonstrate a wide array of fundamental phenomena and technological applications~\cite{mak2016photonics}. Furthermore, their two-dimensional (2D) nature allows for the formation of heterojunctions or homojunctions with arbitrary twist angles resulting in emergent properties~\cite{zhang2021van, alexeev2019resonantly, zhang2020twist, tang2021tuning}. Their strong binding energy and large oscillator strength render them capable of forming polaritons at room temperature \cite{Liu2014}, which has also been shown to retain the intriguing properties of the 2D excitons \cite{Sun2017,Chen2017a,Dufferwiel2017}. Polariton interactions in TMDCs under different configurations are being studied extensively. Realization of Fermi polarons \cite{PhysRevX.10.021011}, trion polaritons \cite{Emmanuele2020, kravtsov2020nonlinear}, and excited-state Rydberg exciton-polaritons \cite{Gu2021} have shown great potential in their abilities to harness strong polariton interactions. Moir\'e exciton-polaritons were recently demonstrated using heterobilayers of WS$_2$/MoSe$_2$ where the electronic confinement potential arising from the twisted heterostructure was shown to enhance the nonlinearity~\cite{zhang2021van}. However, all the platforms above rely on short-range exchange interactions or phase-space filling that pose a bottleneck in realizing few polariton non-linearity under current experimental capabilities. 

In this context, a very attractive possibility is the use of spatially indirect interlayer excitons (IE) in TMDC heterostructures~\cite{rivera2015observation} that can have a permanent dipole moment and hence support highly interacting dipolar polaritons. Formation of polaritons with excitons that possess a permanent dipole moment has been shown to enhance the polariton interactions both in resonant~\cite{PhysRevLett.126.137401,PhysRevLett.121.227402} and non-resonant excitation schemes~\cite{Rosenbergeaat8880}. However, their inherently low oscillator strength creates an impediment in reaching the strong coupling regime without hybridization with a direct intralayer exciton. Furthermore, the large in-built interfacial electric fields make electrical tuning of the energies of the IE far more difficult. Bilayer MoS$_2$ provides a highly attractive platform to solve these issues~\cite{leisgang2020giant, lorchat2021excitons, gerber2019interlayer, peimyoo2021electrical}. The IE in naturally stacked 2H bilayer MoS$_2$ remarkably has both an out-of-plane (static) dipole moment and an in-plane (oscillating) dipole moment. Due to the in-plane dipole moment, the IE in MoS$_2$ bilayer has an oscillator strength of approximately 36\% of that of the intralayer A exciton along with strong absorption that is visible up to room temperatures. Moreover, a strong response to DC electric fields has been demonstrated in these systems, thus confirming their dipolar nature~\cite{leisgang2020giant, lorchat2021excitons, peimyoo2021electrical}. In this work, we achieve a strong coupling of microcavity photons with the IEs (along with intralayer A and B excitons) in bilayer MoS$_2$. The IE polariton shows 10 fold enhancement of the polariton nonlinearity compared to the intralayer A exciton owing to its dipolar and phase space filling contributions. Such enhanced nonlinear response makes them appealing to realize strongly interacting polaritons in condensed matter systems. In addition, the ease of fabrication and realization of multi-polariton species in this system makes it a practical and fundamentally interesting material for studying polariton physics.

\section*{Results}

\begin{figure*}\includegraphics[width=16cm]{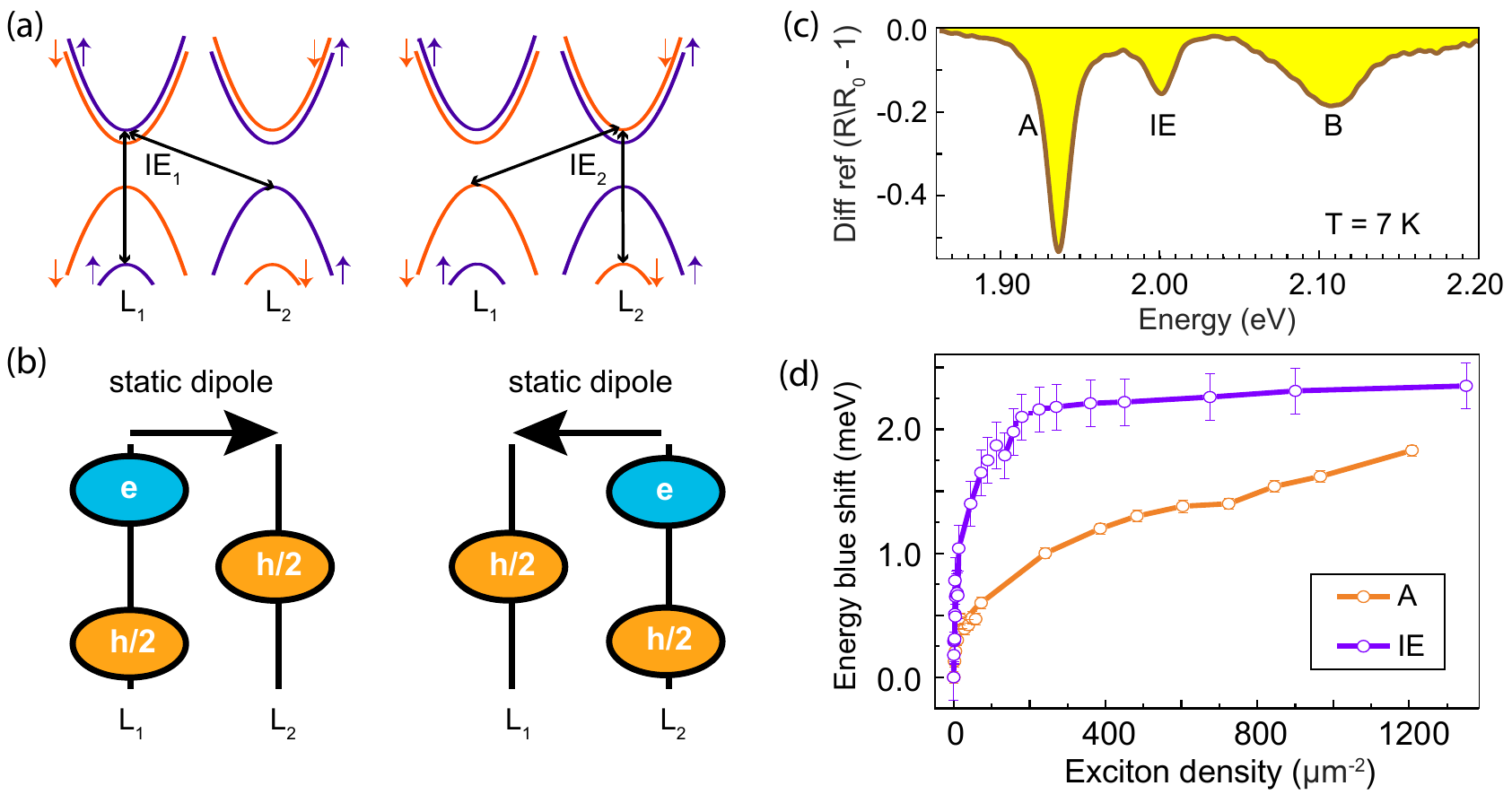}
\caption{ \label{fig:MainFig1} \textbf{Excitons in MoS$_2$ bilayer and white light absorption.} (a) Schematic of the band structure (around the K point) of the bilayer MoS$_2$ showing (black arrows) the participation of the different bands in the IE formation. The blue and red colors represent the up and down spin, respectively. L$_1$ and L$_2$ refer to the layer~1 and layer~2 of the bilayer MoS$_2$, respectively. (b) Schematic showing the two possible charge configurations in the absence of bias voltage between the two layers. In both cases, the electron is localized in one of the layers, and the hole is distributed among both the layers. (c) White-light differential reflection at 7~K showing three dips corresponding to A, IE, and B exciton, respectively. (d) Relative shift of excitonic features observed in instantaneously-probed pulsed reflectivity with increasing exciton density. Pulse duration = 20 ps. The K valley lifetime of A exciton (45~fs) and IE (53~fs) estimated from their respective linewidths are used to estimate the exciton density from the pump fluence. The error bars represent the variance of the fitted parameters caused by noise in the measurement.}
\end{figure*}

Fig.~\ref{fig:MainFig1}a shows a schematic of the bands that form interlayer exciton (IE) in bilayer MoS$_2$. IE$_1$ and IE$_2$ excitons in which electrons are in layer 1, and layer 2 respectively, are energetically degenerate at zero external bias while the hole is delocalized among both layers. The black arrows in Fig.~\ref{fig:MainFig1}a indicate the transitions that form the IE. This peculiar charge distribution is at the heart of producing both in-plane and out-of-plane dipole moments of IE. Fig.~\ref{fig:MainFig1}b shows the charge distribution in real space for IE$_1$ and IE$_2$. Owing to the spatial separation of the electron and the hole, it acquires a permanent dipole moment, the nature of which has been a topic of much interest recently ~\cite{leisgang2020giant, gerber2019interlayer,peimyoo2021electrical,lorchat2021excitons}. IEs in MoS$_2$ homobilayer can be thought of as an admixture of B intralayer exciton with an optically dark but electric field tunable charge-transfer exciton, which is typically found in TMDC heterobilayers~\cite{deilmann2018interlayer}. As a result, IEs in MoS$_2$ homobilayer acquire both a strong oscillator strength and electric field tunability ~\cite{deilmann2018interlayer}. Fig.~\ref{fig:MainFig1}c shows the white light differential reflection of the bilayer sample at zero power limit at 7~K. The peaks at 1.9365~eV, 2.000~eV, and 2.1076~eV correspond to the A, IE, and B excitons, respectively. Interestingly, the absorption of the IE is prominent even at room temperature, which provides a straightforward method to identify the MoS$_2$ bilayer flakes after mechanical exfoliation (see SI Fig.S3).

We probe the nonlinear optical response of the IE and A exciton using resonant pump experiments (see Methods). 
The relative shift of A exciton and IE so obtained are plotted in Fig.~\ref{fig:MainFig1}d against the estimated density of A exciton and IE, respectively (see Methods for the details of the exciton density calculation). Both the excitons blueshift with increased excitation density, with the IE shift being more pronounced. Exchange interaction is the dominant mechanism responsible for the blue shift for the A exciton, while dipolar interactions dominate the interaction in the IE response as exchange interaction is likely to be small given the larger spatial separation of electron and holes~\cite{erkensten2021exciton}. 
To better understand the exciton dynamics, we have also performed ultrafast pump-probe spectroscopy on the bilayers (see SI section \Romannum{10} for the pump-probe data). In this case, the spectra show a redshift of the resonances accompanied by a nearly identical transient response regardless of which exciton is first pumped (A, IE or B). This can be understood if carriers rapidly relax out of the K valley to the band minima on a scale faster than the $\sim$ 200 fs time resolution. This supports our hypothesis that the A and IE transitions are nearly homogeneously broadened with a lifetime dictated by the intervalley scattering rate.  This agrees well with recent coherent spectroscopy and numerical calculations performed for bilayers of MoSe$_2$~\cite{helmrich2021mose2}. The above approach is used to estimate the exciton density in Fig.~\ref{fig:MainFig1}d.

\begin{figure*}
\includegraphics[width=16cm]{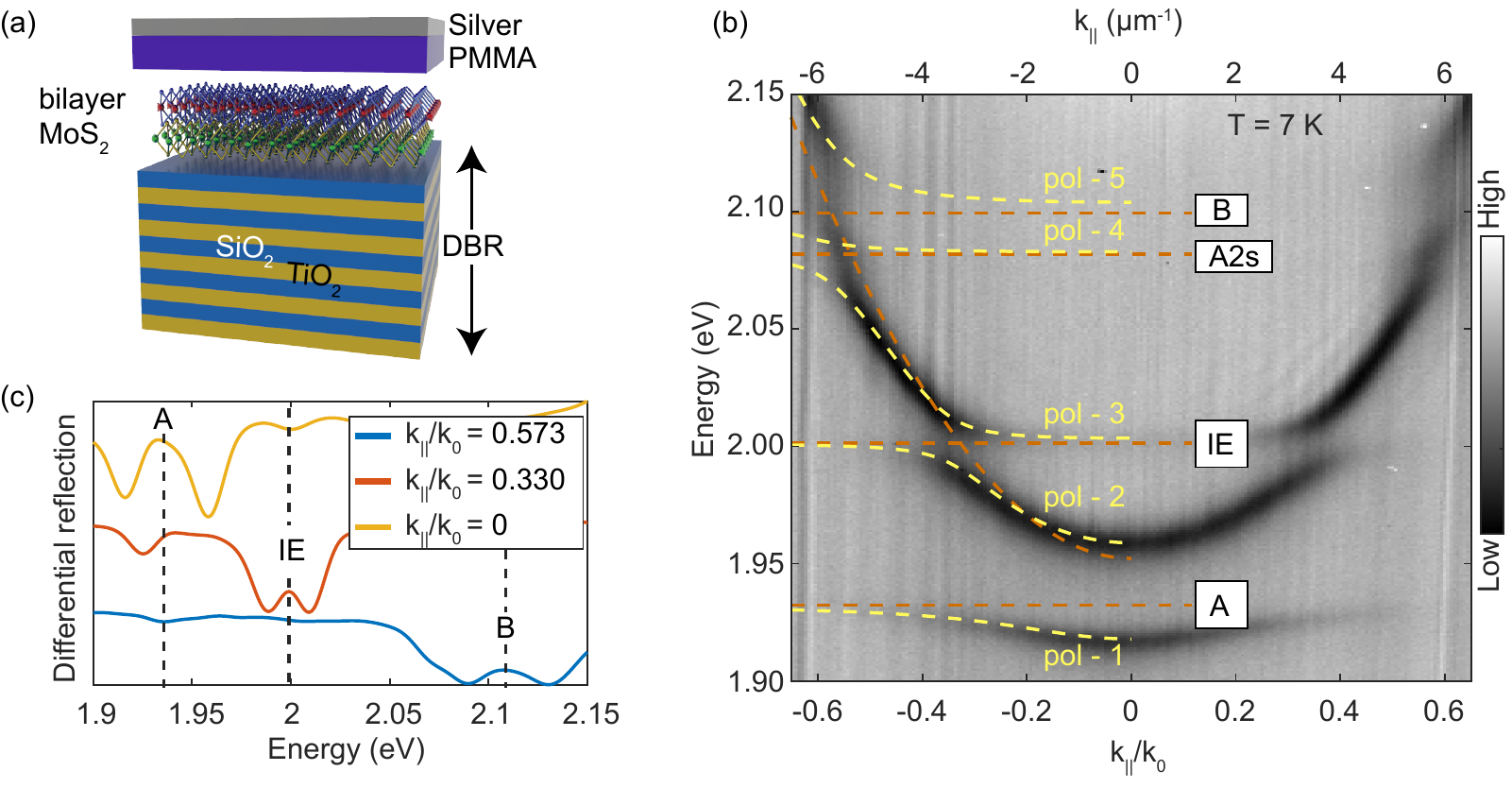}
\caption{ \label{fig:MainFig2} \textbf{Device schematic and polariton dispersion.} (a) Schematic of the device. The bilayer MoS$_2$ is sandwiched between two hexagonal boron nitride flakes of almost similar thickness $\sim$20~nm. The entire stack is transferred by the polypropylene carbonate (PPC) method onto a SiO$_2$ terminated SiO$_2$-TiO$_2$ DBR. As the cavity spacer, a layer of 240~nm thick PMMA was spin-coated on the stack. Finally, 40~nm of Ag was deposited via electron beam evaporation to form a $3\lambda/2$ cavity, where $\lambda$ is the cavity resonance wavelength. (b) White light differential reflection showing all branches of the polaritons at 7~K. 1s and 2s states of A exciton, IE, and B exciton all are strongly coupled to the same cavity mode. This data was measured by illuminating the sample with incoherent halogen light with a power of few nW at the sample.
The yellow dashed lines are the polariton eigenvalues obtained from fit using the five coupled oscillator model. The brown dashed lines denote the energy of the excitons and the cavity mode resulting from the fitting. Here $k_0 = \frac{2 \pi}{\lambda_c}$; $\lambda_c$ is the wavelength of the uncoupled cavity mode at $k_\parallel = 0$. (c) Line cuts at different $k_\parallel$ of the color plot shown in panel-b. It clearly shows that each exciton strongly couples, resulting in two polariton modes that lie above and below the excitonic resonances.}
\end{figure*}

Exciton polaritons were realized in the bilayer MoS$_2$ by embedding it in a microcavity. Fig.~\ref{fig:MainFig2}a shows the schematic of the structure used in this experiment. 
It consists of a bilayer MoS$_2$ encapsulated by thin (20~nm) hexagonal boron nitride (hBN), which is sandwiched between a bottom dielectric distributed Bragg reflector (DBR) mirror and a top silver mirror. See SI Note \Romannum{1} and Methods for the fabrication details and cavity structure. 
The cavity is designed such that the bilayer MoS$_2$ flake sits close to an electric field anti-node, allowing us to observe five dispersive modes associated with the different polaritonic states as shown in Fig.~\ref{fig:MainFig2}b. 
These distinct polariton modes arise due to the hybridization of the cavity photon mode with the various excitonic states present in the bilayer system. We name the polariton branches as pol-1 through pol-5, with pol-1 corresponding to the lowest energy and pol-5 the highest. From the Fig.~\ref{fig:MainFig2}b, we see that the 1s state of A, the IE, B exciton and 2s Rydberg state of A exciton all strongly couple to the same cavity mode at 7~K. 
We fit the data with a five-coupled oscillator model where the energy of all the four excitons, the cavity mode, their Rabi splitting, and the effective refractive index of the system are treated as fit parameters. The fit results in exciton energies $E_{A1s} = 1.9323 \pm 0.0003$~eV, $E_{IE} = 2.0014 \pm 0.0001$~eV, $E_{A2s} = 2.078 \pm 0.002$~eV, $E_{B} = 2.111 \pm 0.001$~eV and the Rabi splittings $\Omega_{A1s} = 40.4 \pm 0.3$~meV, $\Omega_{IL} = 21.4 \pm 0.1$~meV, $\Omega_{A2s} = 13\pm 0.5$~meV, $\Omega_{B} = 51 \pm 0.4$~meV. 
The exciton energies obtained through the fit agree well with the experimentally determined exciton energies with slight shifts due to the strain and changes in the dielectric environment in the cavity. 
In our device, the bare cavity mode is  positively detuned from the A$_\mathrm{1s}$ exciton by $\delta_{C-A_X}$ = 17~meV. 
See SI Note \Romannum{2} for details on the coupled oscillator model. Fig.~\ref{fig:MainFig2}c shows the line cuts of the full k-space data at three different k$_\parallel$ where the cavity mode is resonant with the A exciton (yellow), IE (red) and B exciton (blue), respectively. We notice that the Rabi splitting of the IE polariton remains observable even at 77~K (see SI Fig.S3).

\begin{figure}
\includegraphics[width=8.5cm]{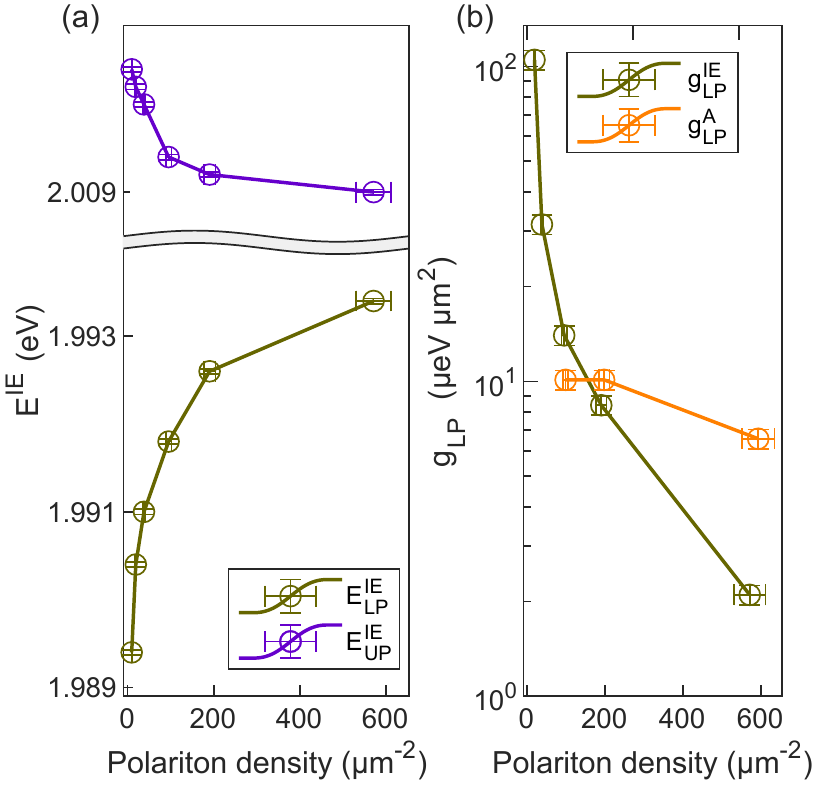}
\centering
\caption{ \label{fig:MainFig3} \textbf{Comparison of nonlinearity of the IE polariton with A exciton polariton.} (a) Energy of the lower and upper branch of the IE polariton as a function of the  polariton density at zero-detuning $k_\parallel$. The movement of the lower branch is more than the upper branch  due to the combined effect of exciton-exciton interaction and saturation. The error bars in energy represent the uncertainty in determining the peak of the Lorentzian fit (polariton energy) to the reflection data. (b) g$^\mathrm{IE}_\mathrm{LP}$ (g$^\mathrm{A}_\mathrm{LP}$) as a function of the IE (A) lower branch polariton density at zero-detuning $k_\parallel$. g$_\mathrm{LP}$ is defined as the local slope of the polariton energy vs. polariton density graph. g$^\mathrm{A}_\mathrm{LP}$ and g$^\mathrm{IE}_\mathrm{LP}$ are calculated from the lower branch of A polariton and IE polariton, respectively. Note that the g$_\mathrm{LP}$ of IE polariton $\sim(100\pm 2)$ \textmu eV \textmu m$^{2}$ is almost 10 times larger than the g$_\mathrm{LP}$ of A exciton polariton $\sim(10\pm 0.2)$ \textmu eV \textmu m$^{2}$. All the power-dependent nonlinear measurements were carried out using a pulsed supercontinuum laser (20 ps pulsewidth) with proper bandpass filter in the input to excite only one polariton species. The polariton density corresponds to the specific branch being excited. The density error bars take into account the error in measuring power and determining the energy of the polariton and linewidth from the Lorentzian fits. The error bars in g$_\mathrm{LP}$ consider the error in determining the density and the energy of the polariton.}
\end{figure}

We measured the white light reflectivity at the IE and A resonance and monitor the energy shifts of the polariton branches, see SI Note \Romannum{5} and SI Note \Romannum{8} for the details of the experiment. Fig.~\ref{fig:MainFig3}a shows the energy of the upper and lower branch of the IE polariton as a function of polariton density at the inplane wavevector corresponding to the zero-detuning $k_\parallel$ (where the cavity mode and exciton energy are degenerate). Two Lorentzian fits at each density is used to obtain the energy of the upper and lower polaritons. SI Note \Romannum{4} discusses the details of the polariton density  calculation and SI Note \Romannum{5} provides the raw data of the density dependent differential reflection of IE polariton. We can see from Fig.~\ref{fig:MainFig3} that the lower branch of the IE polariton moves with pump power more than the upper branch. We also observe a simultaneous increase of the the magnitude of zero-detuning k$_{||}$ along with the reduction of Rabi splitting. This suggests the presence of both exciton energy renormalization and saturation effect. SI Note \Romannum{7} shows the density-dependent Rabi splitting of IE polariton. Taking the derivative of the E$_{LP}$ with respect to density, we calculate the strength of the nonlinearity, g$_{LP}$. Since the rate of blue shift saturates at high density (olive curve in Fig.~\ref{fig:MainFig3}a), g$^{IE}_{LP}$ reduces with polariton density as seen in Fig.~\ref{fig:MainFig3}b. Note that at the lowest powers accessible in our measurement, the IE lower polariton branch already shows a power-dependent blueshift. On the other hand, within the accessible range of the polariton density in our experiment, g$^{A}_{LP}$ remains nearly a constant (see SI Note \Romannum{8} for the blueshift data of the lower polariton branch of A exciton). The obtained strength of the non-linearity in the low density limit for IE and A exciton polaritons are g$\mathrm{^{IE}}_\mathrm{LP}$ $\sim(100\pm 2)$ \textmu eV \textmu m$^{2}$ and g$\mathrm{^{A}}_\mathrm{LP}$ $\sim(10\pm 0.2)$ \textmu eV \textmu m$^{2}$ respectively corresponding to 10 fold increase in the nonlinear response for the IE polaritons.

\section*{Discussion}

We note that the g$\mathrm{^{A}}_\mathrm{LP}$ value 
is comparable to the previous report for A exciton-polariton in monolayer TMDCs \cite{Gu2021,stepanov2021exciton}. 
This allows us to compare the nonlinearity of the IE in bilayer MoS$_2$ with that of Rydberg A$_\mathrm{2s}$ state reported in ref.\cite{Gu2021} -- we find that the nonlinear response of IE is even larger than the A$_\mathrm{2s}$ state by a factor of $\sim$2.7. 
More importantly, we would like to highlight a few critical differences between these two studies. The approximately 4 times enhancement in the g$_\mathrm{LP}$ for A$_\mathrm{2s}$ state in ref.\cite{Gu2021} is attributed to the enhanced phase-space filling due to a higher Bohr radius of 2s compared to 1s. However, this enhancement in nonlinearity comes at the cost of reduced oscillator strength. This is overcome in ref.\cite{Gu2021} by stacking three layers of monolayer TMDC separated by thin hBN. There is, however, no clear experimental route towards using higher-order Rydberg states beyond $n>2s$ in TMDCs to get even higher enhancement as the oscillator strength almost vanishes for higher-order Rydberg states. Additionally, the higher-order Rydberg states are also more susceptible to disorder due to their larger Bohr radius. On the other hand, the spectral signature of interlayer exciton in bilayer MoS$_2$ is visible even at room temperature and was shown to have a significant response to applied out of plane electric field~\cite{leisgang2020giant}. This provides a platform with the possibility of creating highly interacting dipolar polaritons controlled by external electric fields. 

Nonlinear polariton interactions in the present case have contributions from exciton-exciton Coulomb interaction (which includes both dipolar and exchange terms) and phase space filling effect. 
For intralayer excitons with a small Bohr radius only the exchange term and phase space filling effect is important~\cite{erkensten2021exciton}. On the other hand, for interlayer excitons in traditional quantum well systems, the direct dipole-dipole interaction dominates, and the exchange interaction is neglected due to the separation between charges in the two layers~\cite{byrnes2010mott}. In contrast in MoS$_2$ homobilayers, the hole is delocalized across the two layers, and hence there is a finite  overlap with the electron wavefunction leading to small but non-zero exchange interactions. Thus in the present experiments, in addition to the saturation effects, both dipolar and exchange interaction contribute to the overall nonlinear response, with the latter being less significant owing to the finite separation between the electron and hole.

\begin{figure*}
\includegraphics[width=13.8cm]{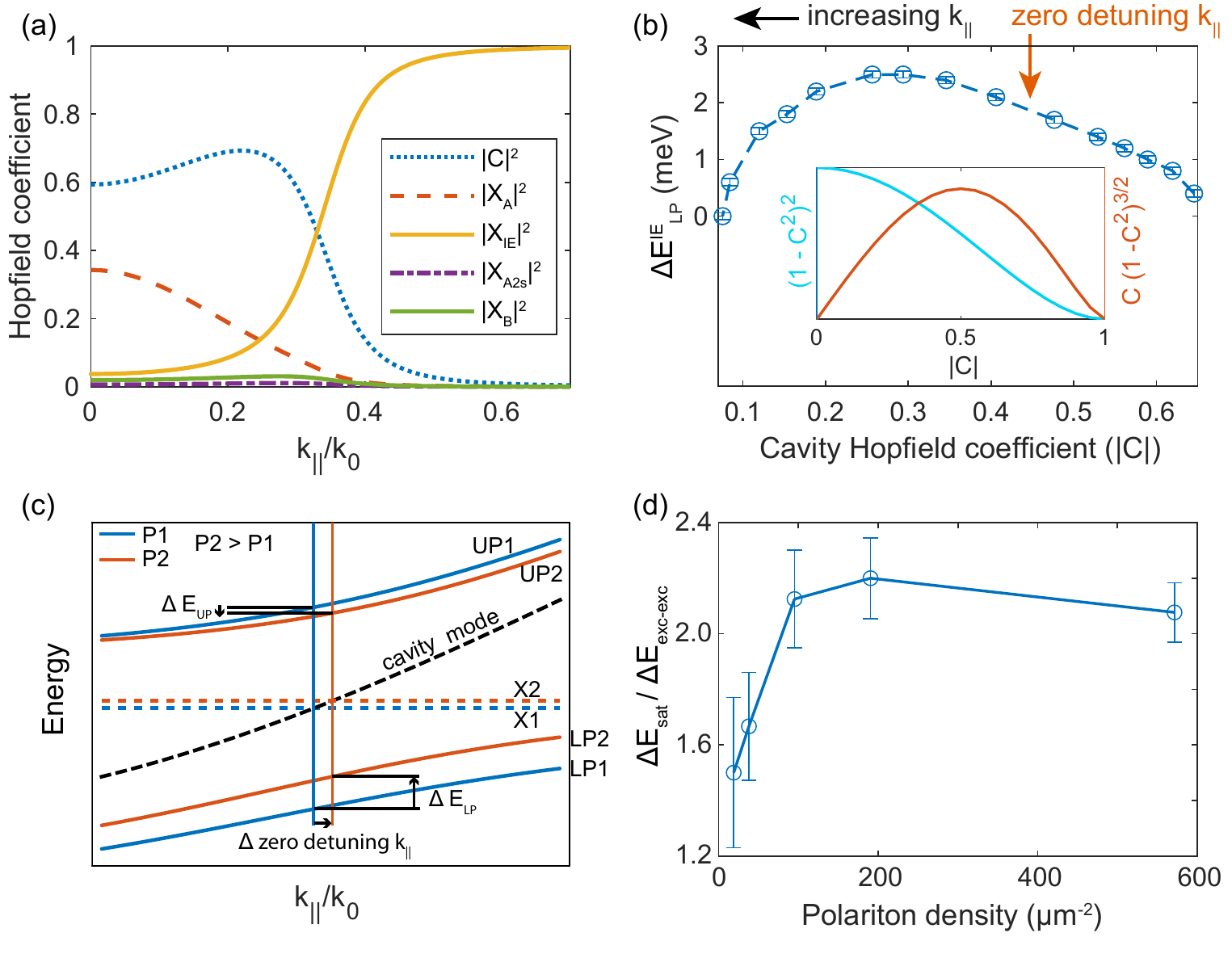}
\caption{ \label{fig:MainFig4} \textbf{Nonlinear response of the IE polariton.} (a) The Hopfield coefficients of the lower IE polariton showing the contributions of cavity photon and the four excitons as a function of $k_\parallel$. (b) Experimentally measured blueshift of the IE lower branch as a function of $k_\parallel$ at a polariton density $\sim 95$ \textmu m$^{-2}$. Here $k_\parallel$ is converted to cavity Hopfield coefficient using the previous plot. The blueshift of the lower branch of IE ($\Delta E^\mathrm{IE}_\mathrm{LP}$) is highest close to the zero-detuning $k_\parallel$, but the shape looks like a tilted parabola. This can be explained by taking both exciton-exciton interaction and saturation effect into account. Inset shows the theoretically expected dependence of exciton-exciton interaction and saturation effect  on $|C|$. It can be  seen that superimposition of these two graphs can explain the tilted parabola shape of $\Delta E^\mathrm{IE}_\mathrm{LP}$ in our data.  (c) Cartoon showing the IE polariton dispersion at two different excitation powers where the combined effect of exciton-exciton interaction and saturation effect is present. Here, X$_1$ and X$_2$ are the exciton energies at power P$_1$ and P$_2$, respectively. The zero-detuning $k_\parallel$ increases with increasing power since the exciton blueshifts. Due to the combined effect of exciton blueshift and saturation, the lower IE polariton branch moves more than upper IE polariton branch. 
(d) Ratio of estimated saturation nonlinearity and exciton-exciton interaction nonlinearity as a function the IE lower branch polariton density. The error bars in energy represent the uncertainty in determining the peak of the Lorentzian fit (polariton energy) to the reflection data.}
\end{figure*}

To discern the contributions to the polaritonic nonlinearity due to dipolar and exchange interaction of excitons ($\Delta\mathrm{E_{exc-exc}}$) versus  phase space filling ($\Delta\mathrm{E_{sat}}$) we carry out the following analysis. Fig.~\ref{fig:MainFig4}a shows the Hopfield coefficients of all the excitonic components and the cavity photon as a function of $k_\parallel$. This is used to obtain the shift in polariton energy, $\Delta E^\mathrm{IE}_\mathrm{LP}$ as a function of Hopfield cavity photon fraction ($C$), see Fig.~\ref{fig:MainFig4}b. The interaction strength due to phase space filling can be written in terms of $g_\mathrm{LP}=4 g_\mathrm{SAT} |C| |X|^3$, where X is the Hopfield coefficient for the exciton. Although this is a multi-exciton system, the contribution of the other excitons close to the zero-detuning $k_\parallel$ of the IE is small. Thus, we can write $g^{IE}_\mathrm{LP}=4 g_\mathrm{SAT} |C| (1-|C|^2)^{3/2}$ which is a non-monotonic function of the cavity Hopfield coefficient as shown in the Fig.~\ref{fig:MainFig4}b inset (red curve). For the case of exchange and dipolar interaction, the interaction strength scales as $g_\mathrm{LP}=g_\mathrm{XX} |X|^4 =g_\mathrm{XX} (1-|C|^2)^2$~\cite{PhysRevLett.121.227402, byrnes2010mott} which monotonically decreases with increasing $|C|$ (cyan curve in Fig.~\ref{fig:MainFig4}b inset). We find that the measured nonlinear response of the IE polariton is the highest at an in-plane wave vector larger than zero-detuning $k_\parallel$, as seen in Fig.~\ref{fig:MainFig4}b. We also notice that $\Delta E ^\mathrm{IE}_\mathrm{LP}$ changes non-monotonically with the cavity Hopfield coefficient. The tilted parabolic shape of the measured $\Delta E^\mathrm{IE}_\mathrm{LP}$ as a function of $|C|$ indicates a roughly equal contributions from both $\Delta\mathrm{E_{exc-exc}}$ and $\Delta\mathrm{E_{sat}}$.

Fig.~\ref{fig:MainFig4}c shows a schematic of the polariton energies at two different powers for $\Delta\mathrm{E_{sat}} > \Delta\mathrm{E_{exc-exc}}$. Due to both the blue shift of the exciton and saturation effect, the lower branch moves with increasing power more than the upper branch. Analyzing the blueshift of the IE lower polariton branch and redshift of the IE upper polariton branch at the zero-detuning $k_\parallel$ we calculate the magnitude of $\Delta\mathrm{E_{exc-exc}}$ and $\Delta\mathrm{E_{sat}}$ in our system. At the zero-detuning $k_\parallel$ we write the energy shift of IE lower branch as $\Delta E^\mathrm{IE}_\mathrm{LP}$ = $\Delta\mathrm{E_{exc-exc}}$ + $\Delta\mathrm{E_{sat}}$ and the energy shift of IE upper branch as $\Delta E^\mathrm{IE}_\mathrm{UP}$ = $\Delta\mathrm{E_{exc-exc}} - \Delta\mathrm{E_{sat}}$. From these equations we obtain $\Delta\mathrm{E_{exc-exc}}$ and $\Delta\mathrm{E_{sat}}$ as a function of polariton density, see SI Note \Romannum{9}. The ratio of $\Delta$E$_\mathrm{sat}$/$\Delta$E$_\mathrm{exc-exc}$ as a function of polariton density is shown in Fig.~\ref{fig:MainFig4}d, indicating that the magnitude of the dipole mediated exciton-exciton interaction nonlinearity and saturation nonlinearity are comparable especially at low polariton densities. Further increase of dipole-dipole interaction by creating imbalance between up and down dipoles (via applied electric field or providing strain) will allow to access regime where dipolar interactions dominate.

We now discuss the origin of the net dipolar interaction arising in our system at the zero electric field.
The two ground states of the interlayer exciton are quasi degenerate at the zero electric field and should not possess net dipole moment in an ideal limit. However, as our experiments indicate, dipolar interactions play a crucial role in the 10 fold enhancement in interaction strength we observe for the interlayer exciton-polaritons. This raises questions about the origin of the dipole moment. Structural asymmetry, strain, residual electric fields, and disorder are the likely reasons for breaking the symmetry and resulting in a net dipole moment. Given that the bilayer MoS$_2$ is sandwiched between two hBN layers of similar thickness, one can rule out the role of structural asymmetry. While local strain and built-in fields could have highly localized effects, we observed similar enhancement over the entire bilayer sample, and hence one can conclude that local strain or fields are also not the key contributors. Finally, if one looks at disorder arising from defects present in the 2D TMDCs, the typical defect densities observed in mechanically exfoliated monolayer MoS$_2$ is on the order of  $10^{12}$-$10^{13}$ $\mathrm{cm}^{-2}$ as has been shown before~\cite{hong2015exploring}. Using the lower bound of $10^{12}$ $\mathrm{cm}^{-2}$ we can estimate the lower bound for dipolar interaction arising from local symmetry breaking between the two layers of the bilayer system. Using a Poisson distribution of defects, we estimate ~ 36\% of the carriers in one layer see a defect in its vicinity, thereby introducing a net dipole moment.  This effect has been hypothesized to be the reason behind the observation of a sizable second harmonic generation in bilayer MoS$_2$ even at zero perpendicular DC electric field~\cite{shree2021interlayer}.  This local imbalance of the up and down dipole moment will increase the dipole-dipole interaction observed in our experiment.

In summary, we demonstrate extremely large optical nonlinearity arising from IE polaritons in homobilayer MoS$_2$. The reported nonlinearity is an order of magnitude larger than that reported for the A excitons in TMDCs, bringing the IE polaritons closer to the polariton blockade regime. The enhanced nonlinear response of the exciton-polaritons is attributed to the dipolar interaction of the interlayer excitons and the saturation nonlinearity with the former starting to dominate in the weak pump limit. 
The homobilayer MoS$_2$ system presents itself as an attractive platform to realize IE polaritons without the need for precise twist angle control while having the potential to reach elevated operational temperatures. 
Using an electric field to induce preferential dipole moment along with lateral photon confinement in the microcavity is expected to push the bilayer MoS$_2$ polariton system into the quantum nonlinear regime.

\clearpage

\section*{Methods}

\section*{Fabrication details}

The DBR used in our experiment is made of 8 pairs of SiO$_2$ (104.9~nm) and TiO$_2$ (64.7~nm) layers and were deposited by radio frequency sputtering on an intrinsic Si chip. MoS$_2$ and hBN were exfoliated from bulk crystals (from 2Dsemiconductor Inc.) using blue tape (Nitto) and scotch tape respectively. MoS$_2$ was exfoliated onto PDMS substrate and hBN was exfoliated onto a 300~nm SiO$_2$/Si substrate. Bilayer MoS$_2$ flakes were identified by reflection spectroscopy -- the additional dip at $\sim$ 639~nm in the reflection signal at room temperature is the hallmark of bilayer MoS$_2$. hBN/bilayer MoS$_2$/hBN heterostructure stacking and transfer were done using the well-known polypropylene carbonate transfer technique~\cite{wang2013one}. The chosen top and bottom hBN layers were of similar thickness $\sim$ 20~nm. The final stack was then transferred onto the DBR at a temperature of 150°C. The chip was kept in chloroform for 12 hours to remove the polypropylene carbonate residue. 240~nm polymethylmethacrylate (PMMA) (495 A4 from Michrochem) was spin-coated to form a 3$\lambda$/2 cavity. Finally, a silver layer (40~nm) was deposited by electron-beam evaporation for the top mirror of the microcavity. Optical images of the device and the bare cavity response are given in SI Note \Romannum{1}.

\section*{Optical measurement details}

We recorded the angle-resolved reflection spectra using the Fourier space imaging technique. A broadband halogen light source was used for reflection measurements. A supercontinuum pulsed light source (NKT Photonics, repetition rate 80~MHz, pulse duration 20~ps) was used to study the power dependence of the bare excitons and polaritons. Appropriate long pass and short pass filters were used in the input to excite a narrow band around the excitons/polaritons under study. For the IE polariton, the excitation bandwidth of the supercontinuum laser was chosen to be $\sim$40 meV around the IE energy in the power-dependent measurements. The set-up was coupled with a Princeton Instruments monochromator with a CCD camera. A 50X, 0.65 numerical aperture objective was used for all the measurements at 7~K in a closed cycle Montana cryostat. The polariton dispersion was revealed by imaging the back aperture of the microscope objective (Fourier plane) onto the camera. The spot size of the laser was 1~\textmu m$^2$ on a uniform area of the sample of dimension $\sim$500~\textmu m$^2$.

\section*{Exciton density calculation in the absence of the top mirror}

For the data presented in Fig.\ref{fig:MainFig1}d, we excite A, IE and B exciton simultaneously at various optical fluences. We have used a supercontinuum laser with a 20 ps pulse width and broadband excitation (1.907~eV to 2.246~eV) scheme for this measurement. At all fluences, each spectrum is fitted with the sum of three Lorentzians to track the evolution of each feature with excitation fluence (see SI Note \Romannum{6}). To calculate the density of individual excitons, we do the following.

We first calculate the absorbed power density by the bilayer MoS$_2$

\begin{equation}
P_\mathrm{absorbed} = \frac{\int_{E_{min}}^{E_{max}}P_{laser}(E) L(E-E_{X}) dE}{A_{beam}}
\end{equation}

Here $P_{laser}(E)$ is the intensity profile of the incident supercontinuum laser, $L(E-E_{X})$ is the absorbance of X exciton around its energy $E_X$, and $A_{beam}$ is the size of the focused laser beam on the sample. We run the integration within 74~meV (53~meV) band around the A exciton (IE) peak energy, respectively. Beyond these bands, the reflection contrast of A exciton and IE are not detectable. The absorbed photon density from the absorbed power density is calculated taking into account the life time of these excitons (45 fs for A exciton 53 fs for IE). We assume all of the absorbed photons form excitons.

\section*{Transient reflectivity measurements}
Transient reflection spectroscopy measurements presented in SI Note \Romannum{10} were obtained using an ultrafast transient absorption spectrometer (Harpia, Light Conversion). An ultrafast optical parametric oscillator (Orpheus-F, Light Converison) provided the pump pulses and white light continuum generation in a sapphire crystal provided the probe pulse. These were filtered using a pinhole inside a telescope and collinearly sent to a home-built microscope equipped with a 50x objective (NA 0.45). The sample was held at 3.5~K using a closed-cycle cryostat (Montana Instruments) overlooked by the microscope. The output lens of the filtering telescope was adjusted to obtain pump and probe spots as large as the bilayer flake. Before its detection by a spectrometer, the pump was removed from the light reflected by the sample using a polarizer so that only the probe beam was detected.

\section*{Error in estimating density of polaritons} 
Currently, the density error bars consider the error in measuring power, determining the energy and linewidth of the polariton. 
However, one primary source of the error is the lifetime of the polaritons. 
To address this issue we have done pump-probe experiment which validates the use of the linewidth of the polaritons to determine the polariton lifetime. Results of our pump-probe experiment is included in the SI note \Romannum{10}.
Another likely source of error is estimating efficient photons' loading into a cavity. This is because the conventional input-output theory does not account for lossy mirrors, as in our experiment. Also photon loading into microcavities from free space is not efficient due to mode mismatch between the free space laser modes and bi-orthonormal modes in lossy cavities~\cite{carusotto2013quantum}. 
However, the relative enhancement of the nonlinearity of IE with respect to A remains unaffected due to the above errors as they exist for both the IE and A exciton polariton.

%\pagebreak %uncomment for page break

\section*{Acknowledgements}
We acknowledge support from ARO through grants W911NF-17-1-0312 and W911NF-22-1-0091; NSF DMR-2011738, and AFOSR-FA2386-21-1-4087. The low temperature spectroscopy facility was supported by NSF MRI grant - DMR-1726573. S.D.L. acknowledges support from a Royal Society Research Fellowship and from the Philip Leverhulme Prize of the Leverhulme Trust.

\section*{Contributions}

V.M.M., M.K. and B.D. conceived the experiments. B.D. fabricated the device. R.B. helped in the fabrication. B.D., M.K. and P.D. performed the measurements. F.T. did the time resolved pump probe measurements. B.D. did the data analysis with inputs from M.K., P.D. and V.M.M. S.D.L. and S.K.C. gave inputs in the data interpretation. B.D, M.K. and V.M.M wrote the manuscript with comments from all authors.

\clearpage

\pagebreak

\pagebreak

\pagebreak

\topmargin=-1.2in

\clearpage

%\input{references_new.bbl}
%\bibliography{MoS2bilayer_ref}

%________________________________________________________________________________________
%Supplementary Material 
%________________________________________________________________________________________
\pagebreak
\begin{center}
\textbf{\large Supplementary Materials: Highly nonlinear dipolar exciton-polaritons in bilayer MoS$_2$}
\end{center}

\setcounter{equation}{0}
\setcounter{figure}{0}
\setcounter{table}{0}
\setcounter{page}{1}

\renewcommand{\theequation}{S\arabic{equation}}
\renewcommand{\thefigure}{S\arabic{figure}}
\renewcommand{\thepage}{S\arabic{page}}
\renewcommand{\bibnumfmt}[1]{[S#1]}
\renewcommand{\citenumfont}[1]{S#1}

\section{Device fabrication}

\begin{figure}[!ht]
\centering
\includegraphics[width=8.5cm]{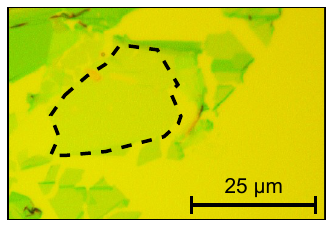}
\caption{ \label{fig:fig1} \textbf{Optical micrograph of the device.} (a) Optical micrograph of the hBN/MoS$_2$ bilayer/hBN heterostructure on a distributed Bragg reflector (DBR) before top silver mirror deposition. The dashed line marks the uniform area of the heterostructure where measurements are done.}
\end{figure}

\begin{figure}[!ht]
\centering
\includegraphics[width=16cm]{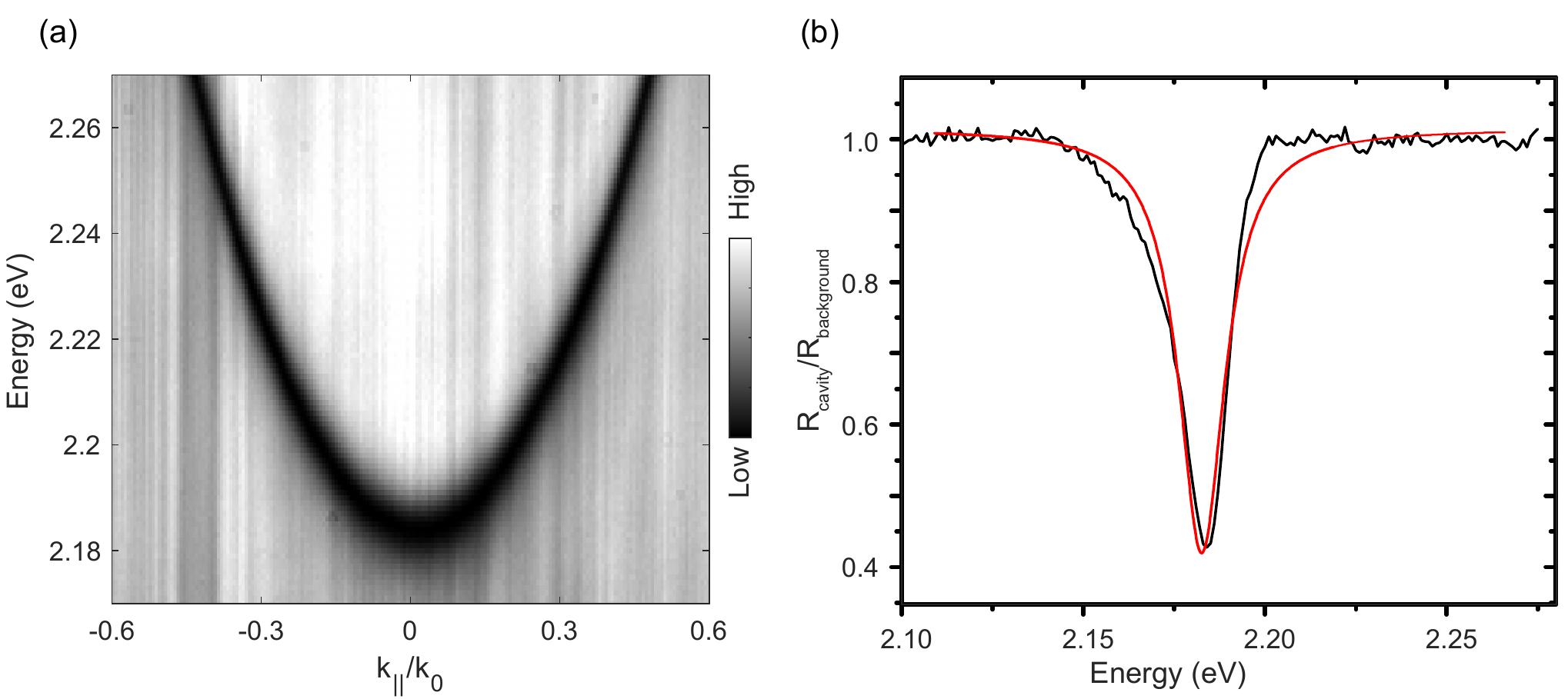}
\caption{ \label{fig:fig2} \textbf{PMMA bare cavity mode.} (a) Dispersion of the PMMA bare cavity mode measured away from the hBN/bilayer MoS2/hBN heterostructure. The bare cavity mode gets 0.23~eV red shifted on the sample area due to the hBN and bilayer MoS$_2$. (b) Line cut of the bare cavity mode at $k_{||} = 0$. The red curve is the fitted Lorentzian with a FWHM of 15.5~meV.}
\end{figure}

\section{Details of coupled oscillator model}

The five polariton branches in our experiment can be described with the eigen modes of the following five coupled oscillator model.

\begin{center}
$\left(
\begin{array}{ccccc}
 \text{$E_{cav}$} & \text{$\Omega _A/2$} & \text{$\Omega _{IE}/2 $} & \text{$\Omega _{A2s}/2$} & \text{$\Omega _B/2$} \\
 \text{$\Omega _A/2$} & \text{$E_A$} & 0 & 0 & 0 \\
 \text{$\Omega _{IE}/2 $} & 0 & \text{$E_{IL}$} & 0 & 0 \\
\text{$\Omega _{A2s}/2$} & 0 & 0 & \text{$E_{A2s}$} & 0 \\
 \text{$\Omega _B/2$} & 0 & 0 & 0 & \text{$E_B$} 
\end{array}
\right)$
\end{center}

where $E_\mathrm{cav} =\frac{c \hbar}{e \mathrm{n_c}} \sqrt{\mathrm{k_v}^2+ k_{||}^2} = \frac{c \hbar}{e \mathrm{n_c}} \sqrt{\mathrm{k_v}^2+\left(10^7\frac{k_{||}}{k_0} \right)^2}$, $k_v = \frac{2 \pi}{\lambda_c}n_c$, $k_0 = \frac{2 \pi}{\lambda_c} \approx 10^7$~nm$^{-1}$, e is the electronic charge.  Here the wavelength is in meter and energy is in eV. The subscripts cav, A, IE, A2s, and B refer to the cavity mode, A exciton, imterlayer exciton, 2s state of A exciton, and B exciton respectively. $\Omega_X$ and $E_X$ refer to the Rabi splitting and energy of the X exciton respectively. 

The fitting yields
$\lambda_c = 637.5 \pm 0.1$~nm, $n_c = 1.450 \pm 0.003$, $E_{A1s} = 1.9323 \pm 0.0003$~eV, $E_{IL} = 2.0014 \pm 0.0001$~eV, $E_{A2s} = 2.078 \pm 0.002$~eV,  $E_{B} = 2.111 \pm 0.001$~eV and the Rabi splittings $\Omega_{A1s} = 40.4 \pm 0.3$~meV, $\Omega_{IL} = 21.4 \pm 0.1$~meV, $\Omega_{A2s} = 13\pm 0.5$~meV,  $\Omega_{B} = 51 \pm 0.4$~meV.

%\clearpage

\section{Reflection measurements on bare exciton and exciton-polariton data at elevated temperature}

\begin{figure}[!ht]
\centering
\includegraphics[width=16cm]{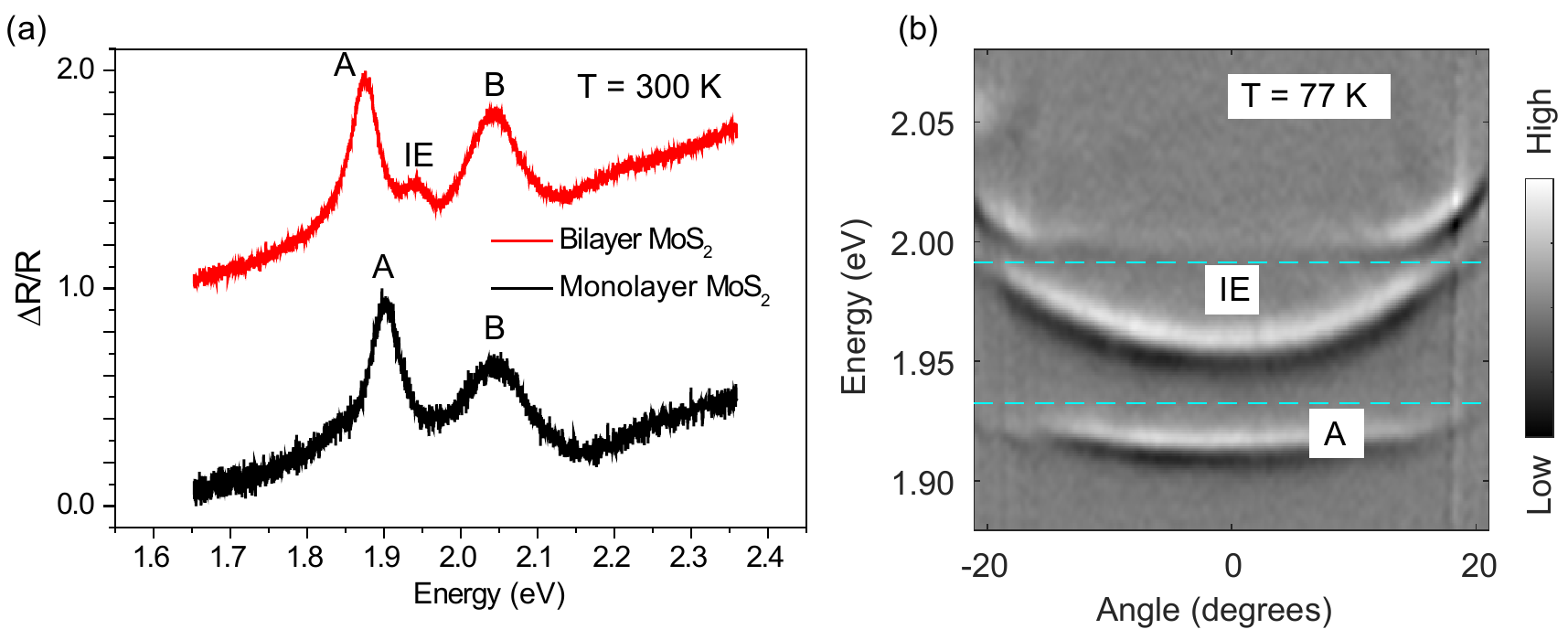}
\caption{ \label{fig:fig3} \textbf{White light reflection  and exciton-polariton at elevated temperatures.} (a) The red (black) color plot shows the differential reflection of bilayer (monolayer) MoS$_2$ on PDMS substrate. The small peak in between the A and B exciton in bilayer is the signature of the IE. (b) Strong coupling of the IE and A exciton at 77~K in bilayer MoS$_2$.}
\end{figure}

\section{Details of the polariton density calculation}

Polariton density is calculated from the Gross-Pitaevskii (GP) equation.
The GP equation at wavevector k can be written as 

\begin{equation}
i \hbar \frac{\partial \psi_{L P}(k, t)}{\partial t}=\left[\epsilon_{L P}(k)-\frac{i \hbar \gamma_{L P}}{2}\right] \psi_{L P}(k, t)+\hbar F_{P}(k, t)
\end{equation}

Here $\epsilon_{L P}(k)=\hbar \omega_{L P}(k)$ is the energy of the lower polariton of an exciton and $\gamma_{LP}$ is the line width of the polariton. $F_{P}(k, t)$ is the coherent driving. The pulse width of our supercontinuum laser is 20 ps which is much larger than the polariton lifetime (30-50~fs). This allows the pumping to be treated as continuous wave which is resonant with our cavity. We can express the pumping term as $F_{P}(k, t)=F_P(k) e^{-i \omega_{P} t}$   and $\psi_{L P}(k, t)=\tilde{\psi}_{L P}(k) e^{-i \omega_{P} t}$ where $\omega_P$ is the pump frequency. This simplifies the above equation to

\begin{equation}
\label{equn2}
\left[\hbar \omega_{P}-\epsilon_{L P}(k)+\frac{i \hbar \gamma_{L P}}{2}\right] \tilde{\psi}_{L P}(k)=\hbar F_{P}(k)
\end{equation}
The $F_{P} (k)$ term can be written from the input-output relation.

\begin{equation}
\label{equn3}
F_{P}(k)=C(k) \sqrt{\frac{\eta \cdot P_{\text {int }}(k)}{\hbar \omega_{P}}}
\end{equation}

\begin{equation}
\label{equn4}
\eta=\frac{\left|t_{\text {top mirror}}\right|^{2}}{\tau_{\text {trip }}}
\end{equation}

Here C(k) is the photon  Hopfield coefficient of the polariton branch. $P_{int}(k)$ is the incident power on the top 
mirror  and $\eta$ is the coupling coefficient. $t_{\text{top mirror}}$
is the transmission of top mirror, and   $\tau_{trip}$
is the photon 
trip time in the cavity. This gives $\eta = 4.1 \times 10^{12}$ s$^{-1}$. Equation~\ref{equn2} can be written as

\begin{equation}
\label{equn5}
\begin{gathered}
\tilde{\psi}_{L P}(k, \omega_{P})=\frac{\hbar F_{P}(k)}{\hbar \omega_{P}-\epsilon_{L P}(k)+\frac{i \hbar \gamma_{L P}}{2}}
\end{gathered}
\end{equation}

Polariton density at the wavevector k and frequency $\omega_{P}$ is

\begin{equation}
\label{equn5_2}
\begin{gathered}
\left|\psi_{L P}(k, \omega_{P}, t)\right|^{2}=\left|\tilde{\psi}_{L P}(k, \omega_{P})\right|^{2}=\frac{|C|^{2} \frac{\eta \cdot P_{\text {int }}(k)}{\hbar \omega_{P}}}{\left(\omega_{P}-\omega_{L P}(k)\right)^{2}+\left(\frac{\gamma_{L P}}{2}\right)^{2}}
\end{gathered}
\end{equation}

Since the polaritons have a finite width in energy, the total polariton density at the wavevector k can be found out by integrating over the all frequency range

\begin{equation}
\label{equn6}
\left|\tilde{\psi}_{L P}(k)\right|^{2}=\int \frac{|C|^{2} \frac{\eta \cdot \xi(k, \omega_{P})}{\hbar \omega_{P}}}{\left(\omega_{P}-\omega_{L P}(k)\right)^{2}+\left(\frac{\gamma_{L P}}{2}\right)^{2}} d \omega_{P}
\end{equation}
Here, $\xi(k, \omega_{P})$ denotes the incident power density (power/frequency/wavevector).

The integral above can be numerically calculated but we can make an approximation to get a close form solution for the density. We note that the Lorentzian part of the integrand for $\gamma_{L P} \rightarrow 0$ is a Dirac delta function 

\begin{equation}
\delta(\omega_P - \omega_{LP})=\lim _{\gamma_{L P} \rightarrow 0} \frac{1}{\pi} \frac{\frac{\gamma_{L P}}{2}}{(\omega_P - \omega_{LP})^{2} + (\frac{\gamma_{L P}}{2})^{2}}
\end{equation}

Using the above formula and taking the power density  $\xi(k, \omega_{P})$ constant over the narrow range of excitation we get

\begin{equation}
\label{equn7}
\left|\tilde{\psi}_{L P}(k)\right|^{2} \approx \frac{2 \pi|C(k)|^{2} \eta \cdot \xi}{\hbar \omega_{L P} (k) \gamma_{L P} (k)}
\end{equation}

Here, $\gamma_{L P} (k)$ is determined from the experimental data. 

Total real space polariton density can be found out by summing $\left|\tilde{\psi}_{L P}(k)\right|^{2}$ in the  k-space for the experimental wavevector range

\begin{equation}
\label{equn7a}
\begin{split}
\left|\tilde{\psi}_{L P}\right|^{2} & = \sum_{k=k_{min}}^{k_{max}} \left|\tilde{\psi}_{L P}(k)\right|^{2} \\
 & = \frac{2 \pi \eta \cdot \xi}{\hbar} \sum_{k=k_{min}}^{k_{max}} \frac{|C(k)|^{2}}{\omega_{L P} (k) \gamma_{L P} (k)}
\end{split}
\end{equation}

If $P_0$ is the measured real space peak power density (power/energy) of the pulsed supercontinuum laser excitation then $\xi = P_0/N$, where N is the number of k point in between $k_{min}$ and $k_{max}$.

The above equation can also be written as 

\begin{equation}
\label{equn7b}
\begin{split}
\left|\tilde{\psi}_{L P}\right|^{2} & = \sum_{k=k_{min}}^{k_{max}} \frac{2 \pi}{\omega_{\mathrm{LP}}(k) \, \tau_{\mathrm{trip}}} \times \frac{ |C(k)|^2 |t_{\mathrm{top \, mirror}}|^2 \, R}{\gamma_{\mathrm{LP}}(k)} 
\end{split}
\end{equation}

Here, $\frac{2 \pi}{\omega_{\mathrm{LP}} \tau_{trip}} \sim 1$ and $R$ is the incident number of photons/sec.

\section{Raw data for the nonlinearity of the IE polariton}

\begin{figure}[!ht]
\centering
\includegraphics[width=16cm]{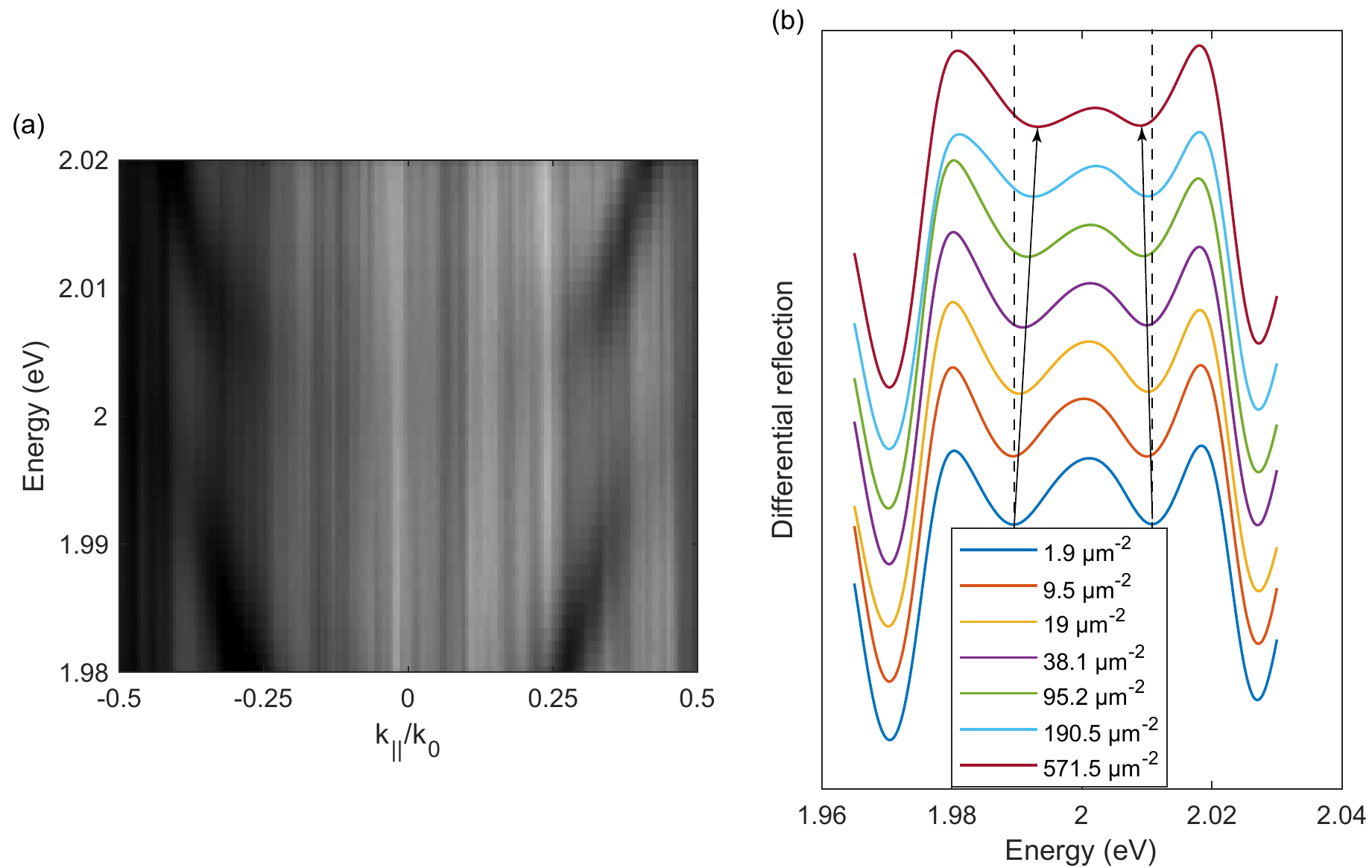}
\caption{ \label{fig:fig4} \textbf{Raw data for the nonlinearity of the IE polariton at 7~K.} (a) K-space differential reflection data showing the dispersion of the IE lower branch (pol-2) and IE upper branch (pol-3) at the lowest density (1.9~\textmu m$^{-2}$). The excitation bandwidth of the supercontinuum laser for the IE is set 40~meV by using a low-pass and a high-pass filter. (b) Line cut of the k-space data at zero detuning $k_\parallel$ showing the differential reflection of the IE lower branch (pol-2) and IE upper branch (pol-3) at different polariton density. Note that since the IE exciton blueshifts with increasing density, the zero detuning $k_\parallel$ also increases with  increasing density. All the power dependent nonlinear measurements (for both samples in the cavity and outside the cavity) were carried out using a pulsed supercontinuum laser (20 ps pulsewidth) with proper bandpass filter in the input to excite only one polariton species.}
\end{figure}

Fig.~\ref{fig:fig4}a shows the differential reflection of the IE polaritons excited (with a 20 ps pulsewidth supercontinuum laser) in a narrow band used to study the nonlinearity of the IE polariton. 
Fig.~\ref{fig:fig4}b shows the differential reflection at the zero detuning $k_\parallel$ for IE polariton. The upper polariton branch red shifts and lower polariton branch blue shifts with increasing density reducing the Rabi splitting.  We also notice that the two branches move asymmetrically -- the lower branch moves more than the upper branch at zero detuning $k_\parallel$. This is because both exciton-exciton interaction and saturation give rise to blue shift for the lower branch but they produce opposite shift for the upper branch -- as described in the main text.

\section{Fits to determine the blueshift of the bare excitons}

\begin{figure}[!ht]
\centering
\includegraphics[width=15.5cm]{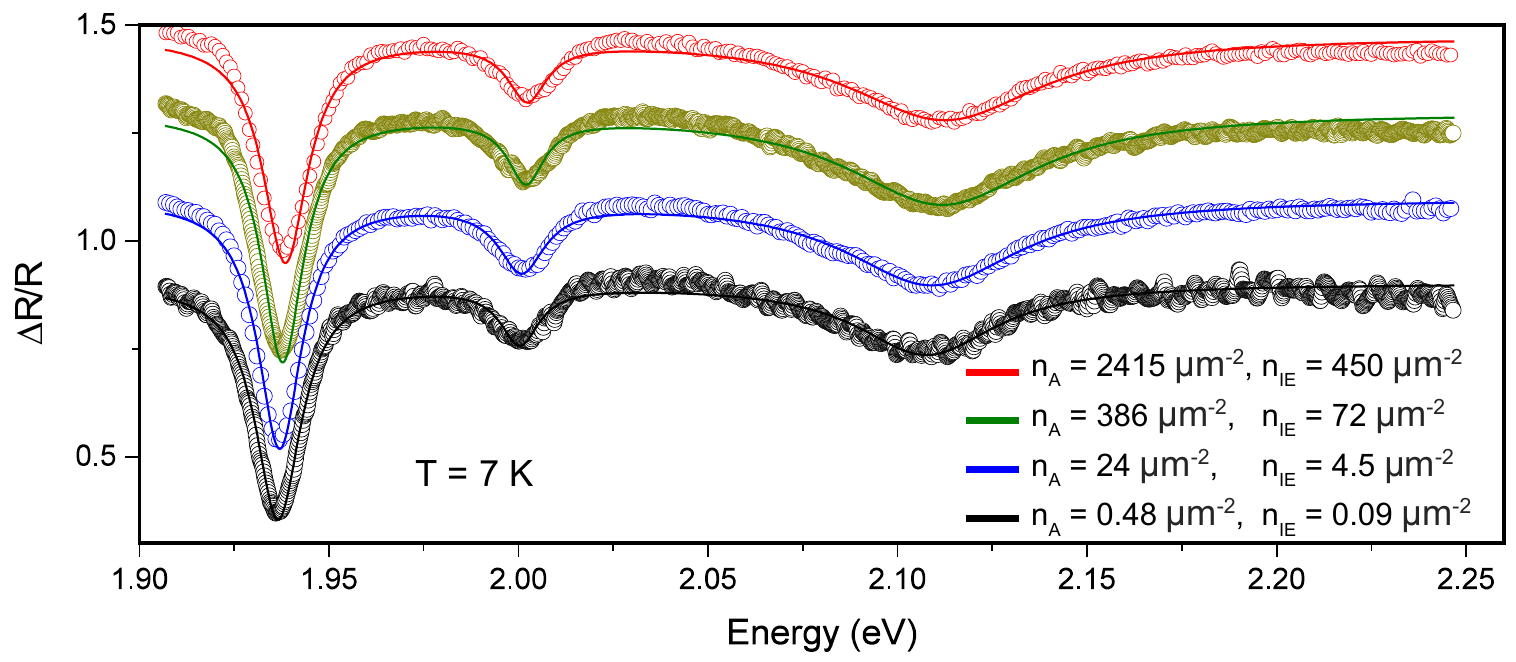}
\caption{ \label{fig:fig6} \textbf{Power dependent reflectance measurement to measure the blue shift. } Three Lorentzian  centered at A exciton, IE and B exciton respectively are fitted to the data at each power. The blue shift of the Lorentzians centered at A exciton and IE are plotted in the main text as a function of estimated exciton density. The method of exciton density estimation is described in Methods.}
\end{figure}
%\clearpage
%\FloatBarrier

\section{Rabi splitting of interlayer exciton polariton as a function of polariton density}

\begin{figure}[!ht]
\centering
\includegraphics[width=8.5cm]{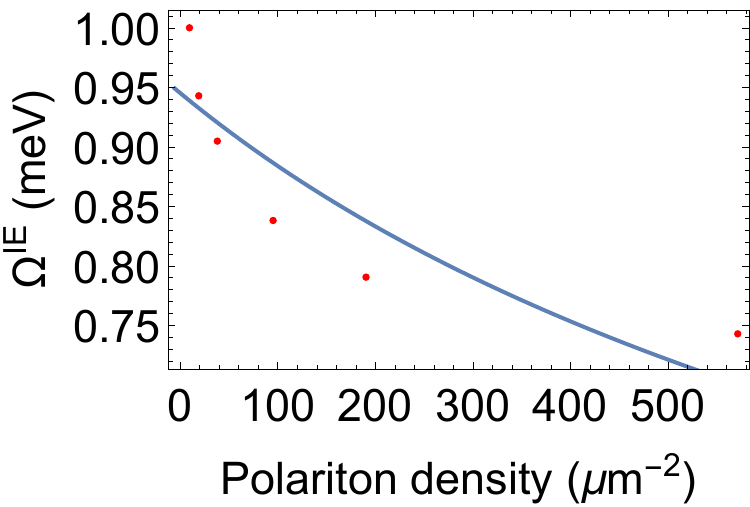}
\caption{ \label{fig:fig8} \textbf{Rabi splitting of IE polariton at the zero detuning $k_\parallel$ as a function of polariton density at 7~K. The blue line is the fit to the usual formula $\Omega =\frac{\Omega _0}{\sqrt{1+\frac{n}{n_{\text{sat}}}}}$ for free 2D excitons.}}
\end{figure}

We plot the Rabi splitting of the IE polariton at the zero detuning $k_\parallel$ as a function of density, see Fig.~\ref{fig:fig8}. We mote that the usual formula for free 
2D excitons $\Omega =\frac{\Omega _0}{\sqrt{1+\frac{n}{n_{\text{sat}}}}}$ does not fit our data well.  The inadequacy of the free 2D exciton formula to describe the density dependence of the Rabi splitting was also observed in a recent study~\cite{zhang2021van1} on the Moir\'e exciton-polariton in a hetrobilayer TMDC.

\section{Energy of the lower branch of A exciton polariton as a function of polariton density}

\begin{figure}[!ht]
\centering
\includegraphics[width=8cm]{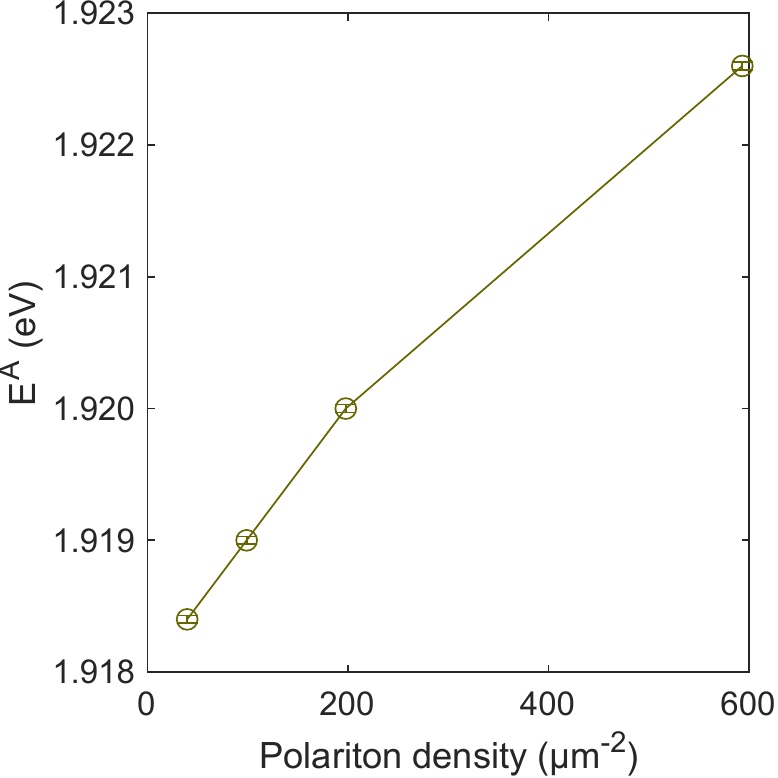}
\caption{ \label{fig:fig9} \textbf{Energy of the lower branch of A exciton polariton as a function of polariton density at 7~K from which the $g_{LP}^A$ is calculated and plotted in Fig.3b of the main manuscript.}}
\end{figure}

\section{Separation of the saturation nonlinearity and exciton-exciton interaction nonlinearity for IE polariton}

\begin{figure*}
\centering
\includegraphics[width=9cm]{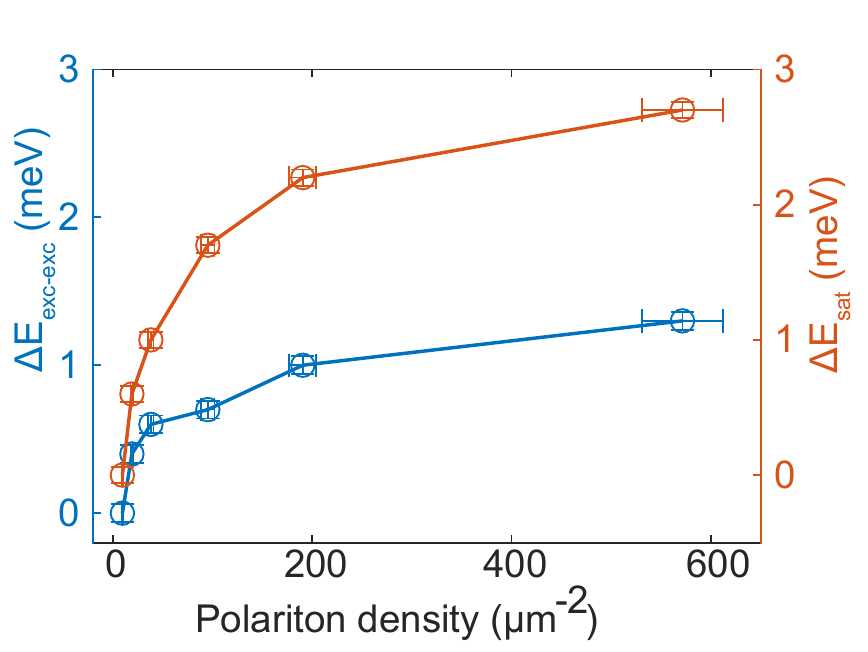}
\caption{ \label{fig:fig10} \textbf{Separation of the saturation nonlinearity and exciton-exciton interaction nonlinearity for IE polariton.}   Calculated $\Delta\mathrm{E_{exc-exc}}$ and $\Delta\mathrm{E_{sat}}$ at the zero detuning $k_\parallel$ from the experimental data as a function polariton density. }
\end{figure*}

\section{Probing non-linearities in the reflectance spectrum of bilayer MoS$_2$ on two distinct time scales}

\begin{figure*}
\centering
\includegraphics[width=16cm]{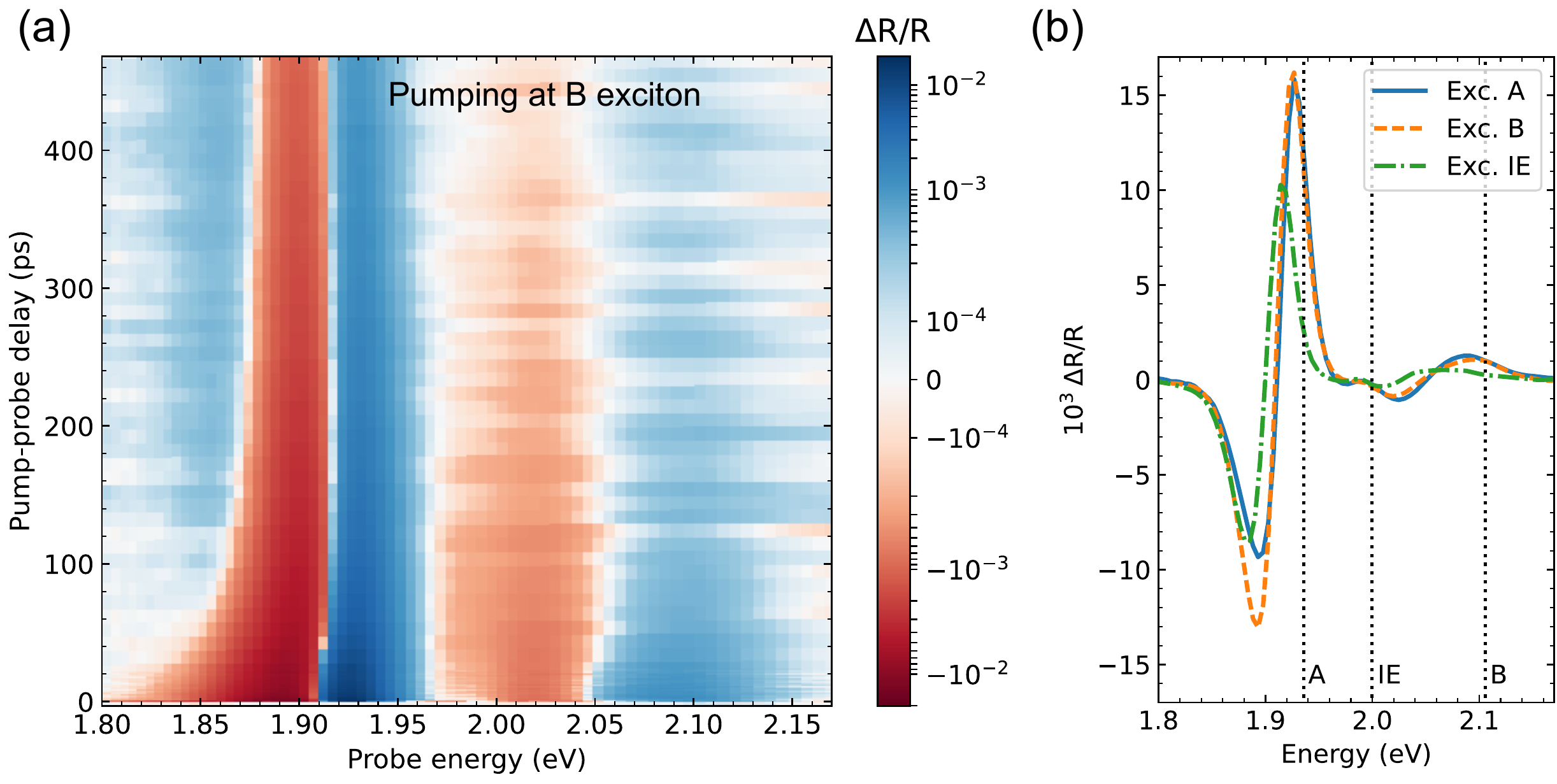}
\caption{ \label{fig:fig13} \textbf{Probing non-linearities in the reflectance spectrum of bilayer MoS$_2$ after the excitons scatter from K-valley.}
(a)  Transient reflectivity spectra probed as a function of white-light probe energy and pump-probe delay. (b) Transient reflectivity spectra probed 1.5~ps after the resonant excitation of  the A (full), B (dashed) and IE (dot-dashed) excitons.   The absorbed photon density is about 4x10$^{4}$ photons/\textmu m$^\textrm{2}$, 2x10$^{4}$ photons/\textmu m$^\textrm{2}$ and 4x10$^{4}$ photons/\textmu m$^\textrm{2}$ respectively.}
\end{figure*}

\begin{figure*}
\includegraphics[width=16cm]{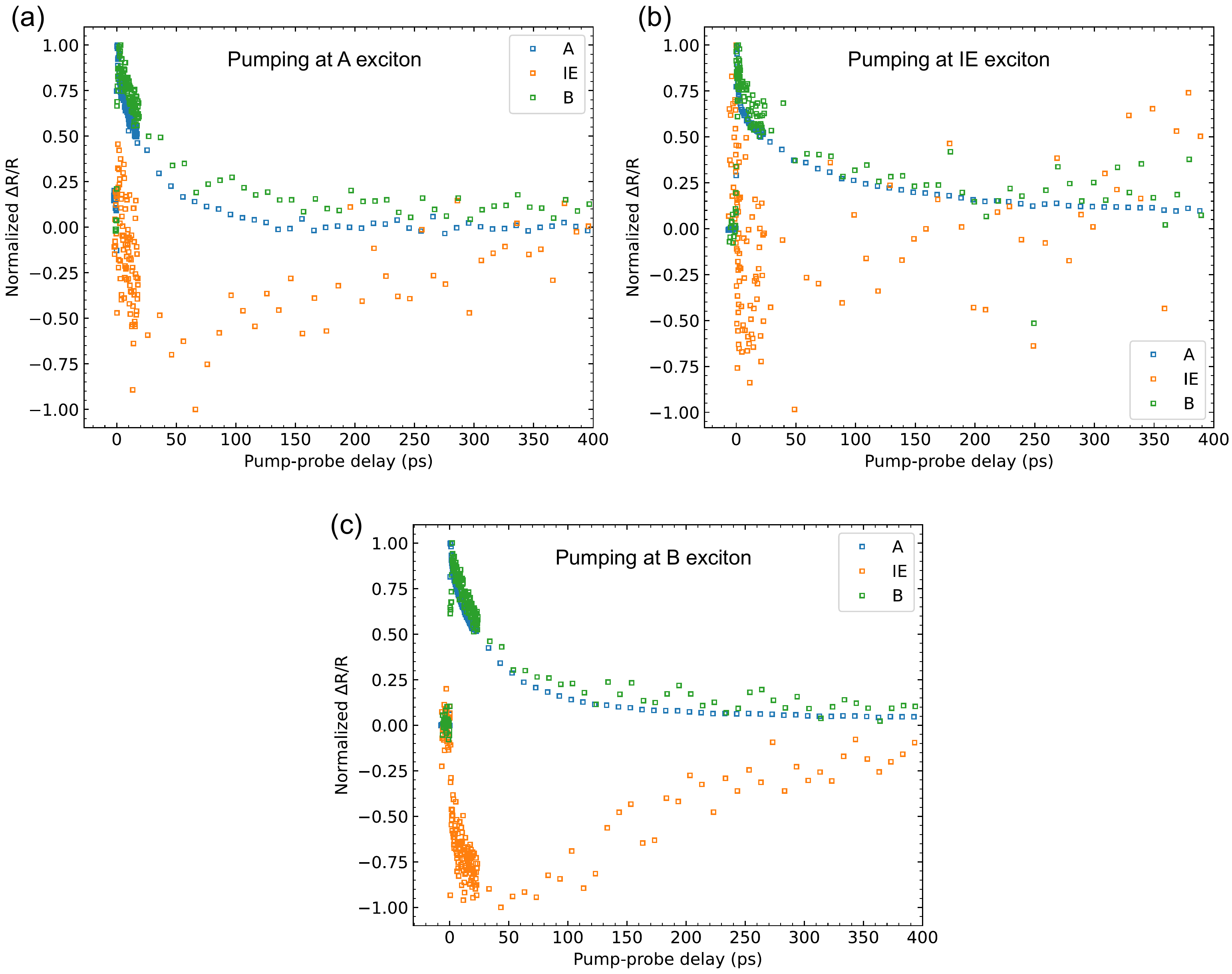}
\centering
\caption{ \label{fig:fig14} \textbf{Probing non-linearities in the reflectance spectrum of bilayer MoS$_2$ after the excitons scatter from K-valley.} Transient reflectivity spectra probed as a function of white-light probe energy and pump-probe delay. The legend corresponds to the probe exciton energy where the time trace is taken.}
\end{figure*}

Fig.~\ref{fig:fig13}a shows transient reflectivity measured on the bilayer film. The sample is first excited by an ultrashort pump pulse (200 fs) resonant with one of the exciton transitions.   The relative change in reflectivity induced by the pump is then probed as a function of time delay using a broadband pulse. The transient reflectivity spectra obtained when pumping the A, B and IE resonances at a pump-probe delay of 1.5~ps are plotted in Fig.~\ref{fig:fig13}b. Regardless of the exciton transition that is initially excited, all the excitonic resonances redshift and saturate (which corresponds to the observed differential lineshapes). Time traces of the transient reflectivity are shown in Fig.~\ref{fig:fig14}.

In contrast, we notice that all the excitonic features blueshift when the sample is probed with a single broadband pulse  lasting 20~ps (main text Fig. 1d). The opposite shifts observed using these two probes provide insight in the excitation regimes induced in the bilayers. When the probe is delayed from the pump  all excitonic features are shifted to lower energies (Fig.~\ref{fig:fig13}a). This is consistent with the expected contribution of relaxed carriers in these materials~\cite{schmidt2016ultrafast1, steinhoff2014influence1, sie2017observation1}. Indeed, the pump-probe delay used in this experiment is much greater than the K-valley exciton lifetime (45 fs for A exciton and and 53 fs for IE). In this case, the probe interacts with the sample when only relaxed carriers remain.

When the reflectivity is probed while the material is still being pumped, as is the case in the single pulse reflectivity experiment, the dominant non-linear effect is instead a shift of all features to higher energies. Indeed, when the pulse interacts with the sample, A, B and IE excitons are continuously created while their constituents quickly relax out of the K valley. This quasi-continuous pumping leads to the steady-state presence of A, IE and B excitons amidst a growing background of relaxed carriers until the pulse no longer interacts with the sample. Moreover, the blueshift arising from excitonic repulsion dominates over the (smaller) redshift from the background of relaxed carriers. Given the long pulses used in our single pulse reflectivity, compared to the lifetime of the excitons ($\sim$ 50 fs), there are about 20~ps/50~fs = 400 times more relaxed carriers than excitons. The strength of the K-valley exciton-exciton interactions are therefore much higher than that due to the interaction with the background carriers.

\clearpage

%\input{references_new.bbl}
%\bibliography{MoS2bilayer_ref}

\end{document}